\documentstyle[a4,12pt,psfig]{article}
\begin{document}

\newcommand{\al}{\alpha}
\newcommand{\bt}{\beta}
\newcommand{\1}{\vspace{1cm}\noindent}
\newcommand{\4}{\vspace{4cm}\noindent}
\newcommand{\be}{\begin{equation}}
\newcommand{\ee}{\end{equation}}
\newcommand{\ba}{\begin{eqnarray}}
\newcommand{\ea}{\end{eqnarray}}
\newcommand{\de}{\delta}
\newcommand{\dd}{\partial}
\newcommand{\ga}{\gamma}
\newcommand{\sg}{\sigma}
\newcommand{\db}{\bar{\partial}}
\newcommand{\ta}{\theta}
\newcommand{\ach}{\mbox{arccosh}}
\newcommand{\bz}{\bar{z}}
\newcommand{\ra}{\rightarrow}
\newcommand{\la}{\lambda}
\newcommand{\nn}{\nonumber}
\newcommand{\p}{\psi^\al(z)}
\newcommand{\pbar}{\bar{\psi}_\beta(z_o)}
\newcommand{\vs}{\vspace{0.2cm}}
\newcommand{\vsa}{\vspace{0.4cm}}
\newcommand{\vsb}{\vspace{0.8cm}}
\newcommand{\half}{\frac{1}{2}}
\newcommand{\eps}{\varepsilon}
\newcommand{\vfi}{\varphi}
\newcommand{\dr}{e^{~~a}_\mu}
\newcommand{\scon}{\omega^{~~a}_\mu}
\newcommand{\pijl}{\leftrightarrow}
\newcommand{\cd}{{\cal D}}
\newcommand{\cb}{{\cal B}}
\newcommand{\cc}{{\cal C}}
\newcommand{\vx}{\vec{x}}
\newcommand{\vp}{\vec{p}}
\newcommand{\om}{\omega}
\newcommand{\ck}{{\cal K}^l_{m,(t)}(\xi,\vfi)}
\newcommand{\cdd}{\cd^l_{m,t}(H,\vfi,\xi)}
\newcommand{\vt}{\vartheta}
\newcommand{\Ml}{{\cal M}^\la_m(\eta,\vt)}
\newcommand{\Mr}{\overline{{\cal M}^{\la}_{m}}(\eta,\vt)}
\newcommand{\Mrs}{\overline{{\cal M}^{\la'}_{m'}}(\eta,\vt)}
\newcommand{\Mrr}{\overline{{\cal M}^{\la}_{m}}(\eta',\vt')}
\newcommand{\ha}{\frac{1}{2}}

\begin{titlepage}
\begin{flushright}
Preprint:THU-97/13\\
hep-th/9706021 \\
June 1997
\end{flushright}
\vsa
\begin{center}
{\large\bf Explicit solutions for Point Particles and Black Holes \vs\\
  in spaces of constant curvature in 2+1-D Gravity\vsa\vsb\\}
           M.Welling\footnote{E-mail: welling@fys.ruu.nl\\
                     Work supported by the European Commission TMR programme
ERBFMRX-CT96-0045}
     \vsa\vsb\\
   {\it Instituut voor Theoretische Fysica\\
     Rijksuniversiteit Utrecht\\
     Princetonplein 5\\
     P.O.\ Box 80006\\
     3508 TA Utrecht\\
     The Netherlands}\vsb\vsa\\
\end{center}
\begin{abstract}
In this paper we consider space-times containing matter expanding or
contracting according to a time-dependent scale factor. Cosmologies with
vanishing, positive or negative cosmological constant are considered. In the
case of vanishing or negative cosmological constant open and closed spatial
surfaces are solutions while in the case of positive cosmological constant only
closed surfaces exist. The gravitational field is solved explicitly in the case
of 1 or 2 particles, 1 black-hole, and 1 black-hole vacuum state.
\end{abstract}
\begin{center}PACS numbers: 0460K, 0420J\end{center}

\end{titlepage}

\section{Introduction}
Shortly after the discovery of explicit solutions for gravitating point
particles in 2+1 dimensions \cite{DJH}, Deser and Jackiw studied point
particles in de Sitter (dS) space-time \cite{DJ}. They found static solutions
where two particles are located at the north and south pole of a sphere. Their
effect on the vacuum solution is to cut out a wedgelike region between two
circles emanating from particle 1 and ending at particle 2. The width of
the wedge is proportional to the particle's mass.

Later Ban\~ados, Teitelboim and Zanelli discovered the possibility of a black
hole in anti de Sitter (AdS) space-time \cite{BTZ}. Subsequent publications
were concerned with the possibility of multi-particle or multi-black hole
states \cite{Steif,Brill,Brill2}. The fact that these multi-object states can
be fully understood is due to the fact that no local degrees of freedom exist
in 2+1-D gravity: all interactions are topological. This fact was used by us
\cite{Welling} and Ciafaloni et. al. \cite{Bellini} to write down explicit
solutions for multi-particle states in 'ordinary 2+1 gravity' ($\Lambda=0$). In
this paper we are concerned with a special class of solutions in 2+1 D
cosmological gravity, namely the ones in which all matter expands or contracts
with the same scale factor. It is typical for gravity in 2+1 dimensions that
the
expansion rate is not influenced by the amount of matter we add to our
space-time. So the scale factor depends on $\Lambda$ and time, but not on the
matter distribution. The geometry of the spatial slice is described by the
Liouville equation (with a different sign for dS and AdS space-time). The
solution to this equation is known in terms of an arbitrary analytic function
$f(z)$. If we add matter to space-time this mapping function becomes
multivalued. For a black hole (BH) we find hyperbolic monodromy, for a black
hole vacuum (BHV) we find parabolic monodromy and for a particle we find
elliptic monodromy. In fact the matter is completely described by these
monodromy transormations (or multivaluedness) of the function $f(z)$. We were
able to find explicit solutions for this function in the case of 1 and 2
particles, of a BH and of a BHV. The problem of finding multi-particle or
multi-BH solutions is equivalent to a Riemann-Hilbert problem with all three
types of monodromy. For vanishing and negative cosmological constant these
solutions were considered on spatially open universes. The question which
matter
distributions form spatially closed
solutions, is answered by the uniformization theorem.

\section{Spaces of constant negative Curvature}
In this section we will describe two space-times that have constant negative
curvature. The simplest example one can think of is to slice ordinary Minkowski
space-time into time surfaces with constant negative curvature:
\ba
X&=&t\sinh\ta\cos\vfi\\
Y&=&t\sinh\ta\sin\vfi\\
T&=&t\cosh\ta
\ea
The Minkowski line element in these coordinates is:
\be
ds^2=-dt^2+t^2(d\ta^2+\sinh^2\ta d\vfi^2)\label{line1}
\ee
The intrinsic curvature of the two dimensional $t=$constant surface is now:
\be
^{(2)}R=-\frac{2}{t^2}
\ee
According to the (contracted) Gauss-Codazzi equations it should be compensated
by the extrinsic curvature $K_{ij}$ as follows:
\be
^{(2)}R-K_{ij}K^{ij}+K^2=0\label{Gauss1}
\ee
where $K$ is defined as the trace of $K_{ij}$. In the above case we have:
\be
K_{ij}=\ha K
g_{ij}~~~~~~~~K=\frac{-1}{\sqrt{g}}\frac{d}{dt}\sqrt{g}=\frac{-2}{t}
\ee
It is important to recognize that we insist on using Gaussian comoving
coordinates defined by:
\be
g_{00}=-1~~~~~~g_{0i}=0
\ee
As it turns out it will be convenient to use complex coordinates in the
following:
\be
z=\tanh(\frac{\ta}{2})e^{i\vfi}~~~~~~\bar{z}=
\tanh(\frac{\ta}{2})e^{-i\vfi}\label{complex}
\ee
The line element becomes:
\be
ds^2=-dt^2+\frac{4S^2(t)}{(1-z\bz)^2}dz~d\bz~~~~~~S(t)=t\label{line2}
\ee
The spatial part is of course precisely the metric on a Poincar\'e disc, with a
scale $S(t)$ depending on time. Up to this point we did nothing but trivial
coordinate transformations. Notice that this metric is invariant under the
following SU(1,1) transformations:
\be
z\ra \frac{\al z+\bt}{\bar{\bt}z+\bar{\al}}~~~~~~~~~~\al,\bt\in~C
\ee
Our next example is Anti-de Sitter space (AdS). In this case we have to include
a negative cosmological constant into Einsteins equations. It is most
convenient to
embed this space into a four dimensional space-time with two negative and two
positive signs:
\be
ds^2=-dU^2-dV^2+dX^2+dY^2
\ee
The constraint determining AdS space-time is given by:
\be
-U^2-V^2+X^2+Y^2=-\ell^2~~~~~~\Lambda=-\frac{1}{\ell^2}
\ee
Both the line element and the constraint are invariant under SO(2,2)
transformations which is the symmetry group of AdS space-time. Because it has
the same dimensionality as the Poincar\'e group, AdS space is also maximally
symmetric. The explicit embedding is given by:
\ba
X&=&\ell\cos\tau\sinh\ta\cos\vfi\\
Y&=&\ell\cos\tau\sinh\ta\sin\vfi\\
U&=&\ell\cos\tau\cosh\ta\\
V&=&\ell\sin\tau
\ea
{}From this embedding we can see that every instant of time AdS space is a
hyperbole with `radius' $\ell\cos\tau$. For $\tau\in (-\ha\pi,0)$ this
hyperbole is contracting, at $\tau=0$ it is momentarily static and for $\tau\in
(0,\ha\pi)$ it is expanding.
Actually, AdS is the universal covering of this space by unwinding the time
coordinate. The line element is:
\be
ds^2=\ell^2\{-d\tau^2+\cos^2\tau(d\ta^2+\sinh^2\ta d\vfi^2)\}
\ee
One immediately recognizes the similarity with (\ref{line1}). Going to complex
coordinates (\ref{complex}) and defining a dimensionful time:
$t=\ell\tau$, gives again the line element (\ref{line2}) but with a different
scale factor:
\be
S(t)=\ell\cos\frac{t}{\ell}
\ee
The intrinsic two dimensional curvature is given by:
\be
^{(2)}R=\frac{-2}{S^2}=\frac{-2}{\ell^2\cos^2\frac{t}{\ell}}
\ee
The Gauss-Codazzi equation now also involves the cosmological constant:
\be
^{(2)}R-K_{ij}K^{ij}+K^2+\frac{2}{\ell^2}=0\label{Gauss2}
\ee
with
\be
K=\frac{2}{\ell}\tan\frac{t}{\ell}
\ee
Notice that at $t=\pm\ha\pi\ell$ both intrinsic and extrinsic curvature diverge
\cite{Brill}.
Notice also that at $t=0$ we have $K=0$. This implies that at $t=0$ the
space-time is momentarily static (or time symmetric)\cite{Steif}.

\section{Inclusion of Point Particles}
In section 1 we studied two vacuum solutions to Einstein's equations with and
without cosmological constant. In this section we will invert the line of
reasoning and {\em assume} a metric of the following form:
\be
ds^2=-dt^2+S^2(t)e^{\phi(z,\bz)}dz d\bz\label{lineelement}
\ee
and solve for $S(t)$ and $e^{\phi}$. Notice that we choose Gauss normal
coordinates (i.e. $g_{00}=-1$ and $g_{0i}=0$). Furthermore we assume that the
gravitational field looks isotropic everywhere. The vacuum solutions of section
1 are found if we set:
\be
e^\phi=\frac{4}{(1-z\bz)^2}
\ee
In the following we will explicitly solve the Einstein equations and find out
whether there is some freedom left to introduce point masses. The Einstein
tensor $G_\mu^\nu$ in Gaussian normal coordinates was calculated by Sachs in
\cite{Sachs}.
With the extra condition of isotropy we find:
\ba
G_i^0&=&0\\
G_0^0&=&-\ha(^{(2)}R+\ha K^2)\\
G_i^j&=&\ha \dot{K}\de_i^j-\frac{1}{4}K^2\de_i^j
\ea
with:
\ba
^{(2)}R&=&-\frac{e^{-\phi}}{S^2}\dd^i\dd_i\phi\label{curvature}\\
K&=&-2\frac{\dot{S}}{S}
\ea
The Einstein equations with cosmological constant are:
\be
G_\mu^\nu+\frac{\de_\mu^\nu}{\ell^2}=8\pi GT_\mu^\nu
\ee
Combining the above formula's we can deduce two relevant equations:
\ba
&&\dd\db\phi-\ha(\dot{S}^2+\frac{S^2}{\ell^2})e^\phi=-4\pi GS^2e^\phi
T^{00}\label{00-eq}\\
&&\ddot{S}+\frac{S}{\ell^2}=0\label{ij-eq}
\ea
where $\dd\equiv\frac{\dd}{\dd z}$, $\db\equiv\frac{\dd}{\dd\bz}$ and $z=x+iy$.
As we are interested in point sources at fixed coordinates $z_n$, we use the
following Ansatz for the energy momentum tensor:
\ba
T^{ij}=T^{0i}&=&0\\
\sqrt{^{(2)}g}T^{00}&=&S^2 e^\phi T^{00}=\sum_n\al_n \de^2(z-z_n)
\ea
where $\al_n$ are constants to be determined later. The next step is to solve
equation (\ref{ij-eq}). In the case where $\Lambda=0$ (or $\ell=\infty$) we
find:
\be
S(t)=At+B
\ee
If the `big bang' takes place at $t=0$ we must set $B=0$. For simplicity we
will normalize the `scaling-velocity' $A=1$. In the case of a non vanishing
cosmological constant we impose the condition that at $t=0$ the solution is
time symmetric: $\dot{S}=0$. The solution is in that case:
\be
S(t)=A\ell\cos(\frac{t}{\ell})
\ee
In the following we will also set $A=1$ for simplicity. Substituting these
solutions into equation (\ref{00-eq}) we find in {\em both} cases:
\be
\dd\db\phi-\ha e^{\phi}=-4\pi G\sum_n\al_n\de^2(z-z_n)
\ee
This equation is the Liouville equation and its general solution in terms of
an analytic and an anti-analytic function is well known:
\be
e^\phi=\frac{4\dd f(z)~\db\bar{f}(\bz)}{(1-f\bar{f})^2}\label{efi}
\ee
If we take $f=z$ and $\bar{f}=\bz$ then we find the vacuum solutions again.
But, as we will see, singularities in the function $f(z)$ (and $\bar{f}(\bz)$)
can be used to produce the $\de$-functions on the right hand side of the
Liouville equation. Notice that we might have expected the more general
solution (\ref{efi}) to the Einstein equations as it is a result of a `residual
gauge transformation' that leaves invariant the line element
(\ref{lineelement}).
If we assume for instance that in the coordinates $f,\bar{f}$ the line element
is:
\be
ds^2=-dt^2+\frac{4}{(1-f\bar{f})^2}dfd\bar{f}
\ee
then after the transformation
\be
f\ra z~~~~~~~~~ \bar{f}\ra\bz\label{residual2}
\ee
the conformal factor $e^\phi$ is precisely given by (\ref{efi}).
The function $f(z)$ however can be {\em multivalued} as we will see. The idea
is very much the same as the ideas employed in \cite{Welling,Bellini} where the
N-particle problem was mapped onto a Riemann-Hilbert problem. In that case we
can always use flat coordinates $u^a$ for which the metric is simply the
Minkowski metric.
But when we include point particles these coordinates turn out to be
multivalued and it becomes convenient to change to single valued coordinates
$z$. The coordinates $u^a(z,\bz)$, viewed as functions of $z$ contain
singularities of the Fuchsian type, causing a branch cut in the $z$-plane.
Because of this branch cut the functions $u^a$ are multivalued. The
singularities also cause the $\de$-functions in the energy-momentum tensor. We
warn the reader that in \cite{Welling,Bellini} the coordinate choice was such
that $K=0$ everywhere.
In the present case this is obviously not true.

\begin{figure}[t]
\centerline{\psfig{figure=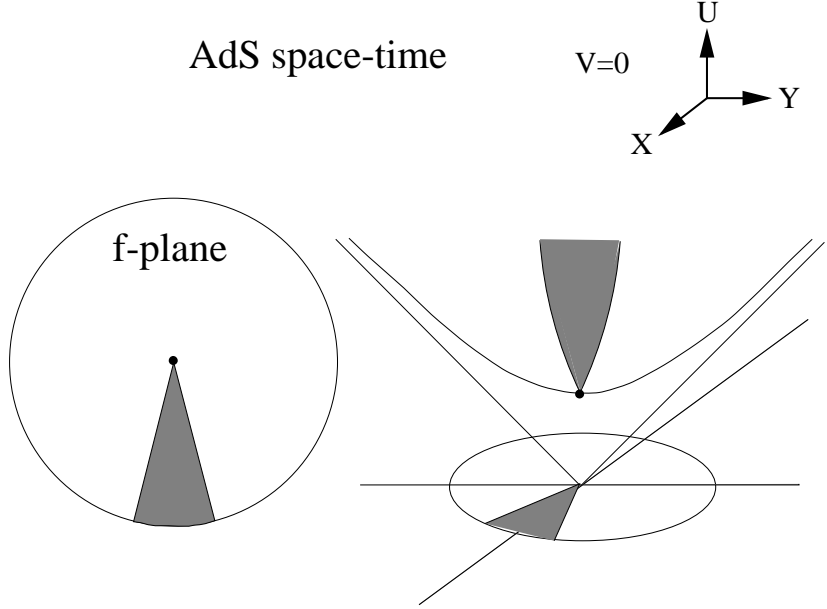,angle=-90,height=8cm}}
\caption{A particle located at the origin in AdS space-time.}
\label{AdS}
\end{figure}

To show that in the present case the singularities are also of the Fuchsian
type let us try to solve the simplest case: one particle of mass $m$ sitting in
the origin. We take a Fuchsian singularity as an Ansatz:
\be
f(z)=(z)^\mu~~~~~~~\bar{f}(\bz)=(\bz)^\mu~~~~~~~~~\mu\in~R
\ee
Substituting this in the Liouville equation gives:
\be
\dd\db(-2\ln(1-(z\bz)^\mu)+(\mu-1)\dd\db\ln(z\bz)
-\ha\frac{\mu^2(z\bz)^{\mu-1}}{(1-(z\bz)^\mu)^2}=
-4\pi Gm\de^2(z)
\ee
The first term cancels with the third term and the second term produces the
$\de$-function by the relation:
\be
\dd\db\ln(z\bz)=\pi\de^2(z)
\ee
The parameter $\mu$ is calculated to be
\be
\mu=1-4Gm
\ee
Notice that the function $f(z)=(z)^\mu$ contains a branch cut due to the
singularity at $z=0$. This cut makes $f$ multivalued:
\be
f(e^{2\pi i}z)=e^{2\pi i\mu}f(z)\label{multi}
\ee
So if the angular coordinate of $z$ runs from 0 to $2\pi$, then the angular
coordinate of $f$ runs from 0 to $2\pi\mu$: see figure (\ref{AdS}).
If we have fixed the number of singularities there is still a residual symmetry
left that can be used to move the singularity around on the Poincar\'e disk:
\be
f\ra g=\frac{\al f+\bt}{\bar{\bt}f+\bar{\al}}~~~~~\in SU(1,1)
\ee
SU(1,1) consists of three basic operations: one rotation and three boosts:
\ba
&R(\vartheta)&:\mbox{Rotation}:\al=e^{i\vartheta}~~~\bt=0\\
&T_x(\xi)&:\mbox{x-Boost}:\al=\cosh\frac{\xi}{2}~~~\bt=\sinh\frac{\xi}{2}\\
&T_y(\xi)&:\mbox{y-Boost}:\al=\cosh\frac{\xi}{2}~~~\bt=i\sinh\frac{\xi}{2}
\ea
On the Poincar\'e disc a boost is really a translation over a distance
$d=\tanh\frac{\xi}{2}$. As we saw, for a particle in the origin the function
$f(z)$ is multivalued (\ref{multi}). If we place a particle at the positive
x-axis at a distance $\tanh\frac{\xi}{2}$ its {\em monodromy} is given by:
\be
g\ra T_x(\xi)R(\vartheta)T^{-1}_x(\xi)g~~~~\vartheta=2\pi(1-4Gm)
\ee
In terms of $\al$ and $\bt$:
\be
T_xRT^{-1}_x:\al=\cos\vartheta+i\sin\vartheta\cosh\xi~~~
\bt=-i\sin\vartheta\sinh\xi
\ee
The next simplest case we can try to solve is the two particle problem.
We can use the SU(1,1) freedom to put one particle  in the origin and the other
at the positive x-axis at a distance $\tanh\frac{\xi}{2}$ from the origin. But
we know the monodromy of the function $f(z)$:
\ba
f&\ra& R(2\pi\mu_1)f~~~~~~\mbox{at}~f=0\\
f&\ra& T_x(\xi) R(2\pi\mu_2)T^{-1}_x(\xi)f\label{mon2}~~~~~~
\mbox{at}~f=\tanh\frac{\xi}{2}
\ea
with $\mu_n=1-4Gm_n$. The fact that we try to solve the function $f(z)$ from
the information contained in its monodromies is called the Riemann-Hilbert
problem. It is deeply related to the theory of Fuchsian differential equations.
One can show that $f(z)$ can be written as $\frac{y_1}{y_2}$ where $y_i(z)$ are
the two independent solutions to the following differential equation
(\cite{Bieberbach,Hille,Bellini}):
\be
y''+\frac{1}{4}(\frac{1-\mu_1^2}{z^2}+\frac{1-\mu_2^2}{(z-1)^2}+
\frac{1-\mu_1^2-\mu_2^2+\mu_\infty^2}{z(z-1)})y=0
\ee
At the points $z=0,1,\infty$ this differential equation is singular, but the
singularities are of a special type called Fuchsian singularities. In a local
basis where the monodromy is diagonal at $z=z_n$ the singular behaviour of the
two independent solutions is:
\ba
y_1&\stackrel{z\ra z_{1,2}}{\sim}&(z-z_{1,2})^{\ha(1+\mu_{1,2})}~~~~~
y_1\stackrel{z\ra\infty}{\sim}(\frac{1}{z})^{-\ha(1+\mu_{\infty})}
\label{local1}\\
y_2&\stackrel{z\ra z_{1,2}}{\sim}&(z-z_{1,2})^{\ha(1-\mu_{1,2})}~~~~~
y_2\stackrel{z\ra\infty}{\sim}(\frac{1}{z})^
{-\ha(1-\mu_{\infty})}\label{local2}
\ea
with $z_1=0$ and $z_2=1$
In order to produce the correct $\de$-functions we want this singular behaviour
locally to be the same as the behaviour we found in the case of one particle,
so we may write:
\be
\mu_{1,2}=1-4Gm_{1,2}~~~~~~\mu_\infty=1-4GH_{\bf tot}
\ee
where $H_{\bf tot}$ is the total energy of the system in the center of mass
frame defined by:
\be
\cos(4\pi GH_{\bf tot})=\cos(4\pi Gm_1)\cos(4\pi Gm_2)-\cosh{\xi}
\sin(4\pi Gm_1)\sin(4\pi Gm_2)\label{totalenergy}
\ee
If we use a coordinate system for which the monodromy for particle 1 is
diagonal, the solution around particle 2 and around infinity is, by analytic
continuation of the solution around particle 1, a linear combination of the
$y_i$ defined in (\ref{local1},\ref{local2}). As we saw, the solution to the
Liouville equation is $f=\frac{y_1}{y_2}$. As the Liouville equation is
nonlinear, the relative constant between $y_1$ and $y_2$ is important. This
constant is determined by matching the monodromy (\ref{mon2}) to the monodromy
of $f=\frac{y_1}{y_2}$. Once this is done we find as a solution:
\be
f(z)=z^{\mu_1}\coth\frac{\xi}{2}\frac{\tilde{F}[\ha(1+\mu_\infty+\mu_1-\mu_2),
\ha(1-\mu_\infty+\mu_1-\mu_2),1+\mu_1;z]}{\tilde{F}
[\ha(1+\mu_\infty-\mu_1-\mu_2),\ha(1-\mu_\infty-\mu_1-\mu_2)
,1-\mu_1;z]}\label{Fuchs}
\ee
where:
\be
\tilde{F}[a,b,c;z]=\frac{\Gamma(a)\Gamma(b)}{\Gamma(c)}F[a,b,c;z]
\ee
and $F$ denotes the hypergeometric function.
Because the hypergeometric function has its singularities at $z=0,1,\infty$ the
particles are located in these coordinates at $z=0,1$. They can however easily
be moved to arbitrary locations by using the following transformation:
\be
w=a_1+z(a_2-a_1)
\ee
In the $f$-coordinate system the first particle is located at $|f|=0$. This
fact can be checked by calculating $f(0)=0$ from (\ref{Fuchs})
\footnote{$F(a,b,c;0)=1$}. We also deduced that the second particle was located
at the fixed point of the SU(1,1) monodromy, i.e. at
$x=\tanh\frac{\xi}{2},y=0$. If we use the analytic continuation of $F(a,b,c;z)$
in the neighbourhood of $z=1$ and evaluate $f(1)$ we indeed find:
\be
f(1)=\tanh\frac{\xi}{2}
\ee
It is of course interesting to see where $f(\infty)$ is located in the
$f$-plane. For this one uses the analytic continuation of $F(a,b,c;z)$ around
infinity (or $w=\frac{1}{z}\ra 0$) and one finds (see for instance
\cite{complex}, page 663):
\be
f(\infty)=e^{4\pi i Gm_1}\sqrt{\frac{\sin\{2\pi G(H+m_1+m_2)\}
\sin\{2\pi G(H+m_1-m_2)\}}{\sin\{2\pi G(H-m_1-m_2)\}\sin\{2\pi G(H-m_1+m_2)\}}}
\ee
Notice that in the range $m_1,m_2,H\in [0,\frac{1}{4G}]$ we have
$|f(\infty)|\geq 1$. But physical infinity is already at $|f|=1$, implying that
our universe is open! If we take the limit $m_1\ra H,~m_2\ra 0,~\xi\ra 0$ we
find that $|f(\infty)|\ra\infty$ which is the correct result for the one
particle case.

Equation (\ref{totalenergy}) is precisely equation (4.5) in \cite{Steif} for
the case of two particles in an open universe without a horizon. Steif finds
that for different ranges of the masses $m_i$ and $\xi$ there are also
solutions representing an open universe where the two particles are enclosed by
a horizon and a closed universe with an additional image mass. It would be
interesting
to see whether our gauge supports these solutions too.

If we want to insert N particles on the Poincar\'e disc we need to solve the
Riemann-Hilbert problem with N Fuchsian singularities. One line of attack is to
 write down a more general second order differential equation of the form:
\cite{Bellini,Hille,Bieberbach}:
\be
y''+\frac{1}{4}\sum_i(\frac{1-\mu_i^2}{(z-z_i)^2}+\frac{2\bt_i}{(z-z_i)})y=0
\ee
The $\mu_i$ are again used to match the masses and the $\bt_i$, called
accessory parameters, are to be determined by matching with the monodromy. The
problem of finding the explicit solution is however frustrated by the
appearance of apparent singularities. These are singularities of the
differential equation but zero's of the solution to that equation.

Another way to think about the general Riemann-Hilbert problem is to try to
write down a first order matrix differential equation \cite{Sato,Welling}:
\be
\dd Y=\sum_{i=1}^N\frac{A_i}{z-z_i}Y
\ee
where $A_i$ are constant matrices (possibly depending on the $a_i$) and $Y$ is
a fundamental matrix of solutions. The Riemann-Hilbert problem is now
equivalent to finding the matrices $A_i$, given the monodromy matrices and the
local exponents at the particle singularities. Lappo-Danilevski was the first
to solve this problem in terms of an infinite series of {\em hyperlogarithms}
\cite{Chud}. Although this proves that under certain reasonable assumptions
there exist solutions, these solutions are of little practical value because of
their enormous complexity.

\section{Black Hole solution}
\begin{figure}[t]
\centerline{\psfig{figure=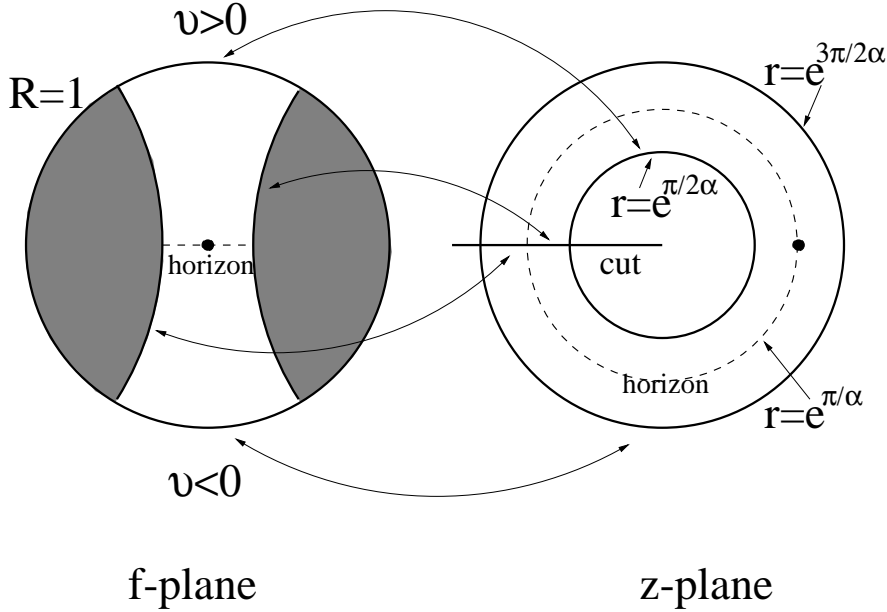,angle=-90,height=8cm}}
\caption{The mapping function $f(z)$ for a black hole.}
\label{annulus}
\end{figure}

Up till now we only considered  point particles, but it is well known that in
AdS
space-time also black hole (BH) solutions exist \cite{BTZ, Steif, Brill}.
In the case of particles we were concerned with a  multivalued mapping function
$f(z)$ having elliptic monodromy. But SU(1,1) also contains hyperbolic and
parabolic transformations. It happens that the black hole of mass $M$ is
characterized by the following hyperbolic monodromy:
\be
f(z)\ra \frac{\cosh\pi(\sqrt{8GM})~f(z)+\sinh\pi(\sqrt{8GM})}
{\sinh\pi(\sqrt{8GM})~f(z)+\cosh\pi(\sqrt{8GM})}\label{hyp}
\ee
We could of course also have chosen a rotated transformation to change the
orientation of the BH. We could also easily move the BH to a different location
by conjugating (\ref{hyp}) by a boost. Our task is thus to find the multivalued
function $f(z)$ with monodromy (\ref{hyp}). For that we first diagonalize the
SU(1,1) transformation\footnote{In this paper we do not distinguish between
SU(1,1) and its projected counterpart PSU(1,1). By `diagonalizing' we imply
that we diagonalize SU(1,1)}.
\be
f'\ra\frac{\cosh\pi(\sqrt{8GM})-\sinh\pi(\sqrt{8GM})}
{\cosh\pi(\sqrt{8GM})+\sinh\pi(\sqrt{8GM})}f'
\ee
where $f'$ is $f$ in a diagonal basis. Next consider the function:
\be
f'=z^{i\al}~~~~~~~~\al\in R
\ee
After a full rotation we find:
\be
f'(e^{2\pi i})z=e^{-2\pi\al}f'(z)
\ee
So if we take:
\be
\al=\frac{-1}{2\pi}\ln(\frac{\cosh\pi(\sqrt{8GM})
-\sinh\pi(\sqrt{8GM})}{\cosh\pi(\sqrt{8GM})+
\sinh\pi(\sqrt{8GM})})=\sqrt{8MG}
\ee
then $f'(z)$ has the correct monodromy. Transforming back to the old basis
finally gives:
\be
f(z)=\frac{1+z^{i\sqrt{8GM}}}{1-z^{i\sqrt{8GM}}}=i\cot(\sqrt{2GM}\ln
z)\label{mapf}
\ee
Next we shall study this function a bit more. First we invert relation
(\ref{mapf}) to:
\be
z^{i\al}=\frac{f-1}{f+1}~~~~~~~\al=\sqrt{8GM}
\ee
Putting $z=re^{i\vfi}$ and $f=Re^{i\vt}$ we write:
\be
e^{-\al\vfi}(\cos(\al\ln r)+i\sin(\al\ln r)=\frac{R^2-1}{R^2+1+2R\cos\vt}+
i\frac{2R\sin\vt}{R^2+1+2R\cos\vt}
\ee
Taking $\vt=0,\pi$ (the real line on the f-disc) we get:
\be
\vt=0,\pi:~~~r=e^{\frac{\pi}{\al}(2n+1)}~~~\vfi=
\pm\frac{1}{\al}\ln(\frac{1+R}{1-R})
\ee
The conformal factor diverges at the edge of the disc: $R=1$. There we find:
\be
R=1:~~~r=e^{\frac{\pi}{2\al}(2n+1)}~~~\vfi=
\frac{-1}{\al}\ln(\tan\frac{|\vt|}{2})
\ee
where $\vt>0$ implies even values of $n$ and $\vt<0$ implies odd values of $n$.
Putting this information together we find figure (\ref{annulus}).
On the $f$-plane one recognizes the BH geometry. The dotted line is the horizon
(at $t=0$) Note that the location  of infinity ($R=1$) in the interior region
and exterior region of the BH are mapped to different circles on the z-plane.
For a more detailed description of this BH solution represented on the
Poincar\'e disc we refer to \cite{Brill}.

\begin{figure}[t]
\centerline{\psfig{figure=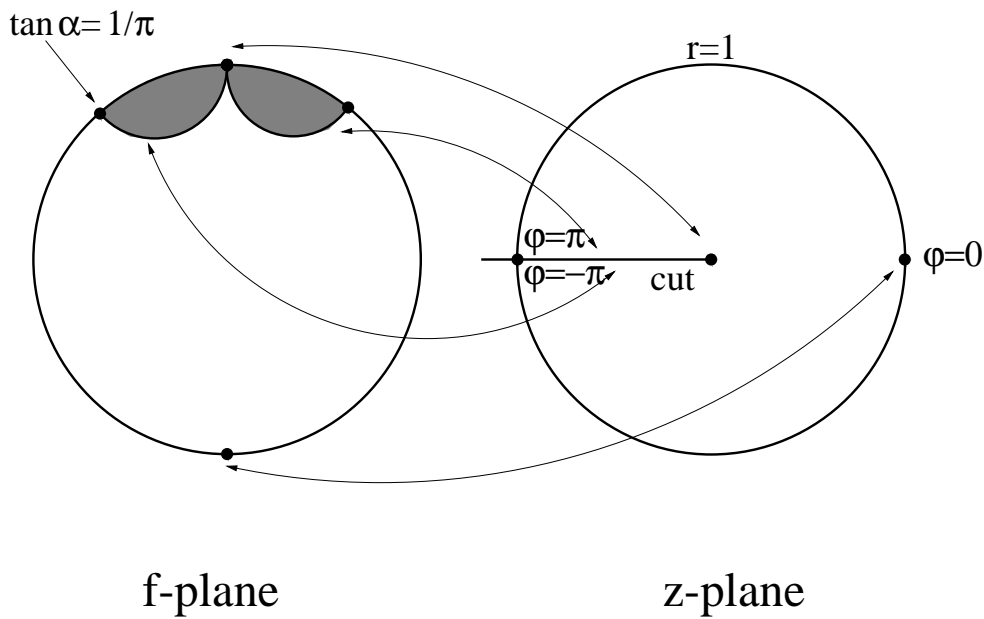,angle=-90,height=8cm}}
\caption{The mapping function $f(z)$ for a black hole vacuum solution.}
\label{bhv}
\end{figure}

So we find that elliptic monodromy for the mapping function $f(z)$ gives a
particle and hyperbolic monodromy a black hole. But what about the parabolic
monodromy? In for instance \cite{Steif} we find that this represents the black
hole vacuum solution. One can either view this as the limit of a BH with zero
mass or a particle with mass $1/4G$. It represents an infinitely long,
and infinitely thin throat. The monodromy is given by:
\be
f(z)\ra \frac{(1+i\pi)f+\pi}{\pi f+(1-i\pi)}
\ee
The mapping function that exhibits this monodromy can be found by transforming
this
monodromy to Jordan form:
\be
f'(z)\ra f'(z)+2\pi i
\ee
and recognize the logarithm. Transforming back gives:
\be
f(z)=i\frac{\ln z+1}{\ln z-1}\label{mapf2}
\ee
This mapping is depicted in figure (\ref{bhv}).
The next step would be to search for mapping functions that describe multi BH
states or combinations of BHs and particles. One should try to find mapping
functions that exhibit prescribed monodromy properties. Only in this case we
may have combinations of elliptic, hyperbolic and parabolic transformations!
One could view this as a kind of generalized Riemann-Hilbert problem. Needless
to say that the more BHs and particles one wants to describe the more
complicated
$f(z)$ gets. For the case of two `objects' the equation that relates the masses
is analogous to (\ref{totalenergy}). For instance, in the case of two BHs we
have \cite{Brill2}:
\ba
\cosh\pi(\sqrt{8GM_{\bf
tot}})=&&-\cosh\pi(\sqrt{8GM_1})\cosh\pi(\sqrt{8GM_2})\\
&&+\cosh(\frac{d}{\ell})\sinh\pi(\sqrt{8GM_1})\sinh\pi(\sqrt{8GM_2})
\ea
where d is the distance between the horizons of the BHs. One recognizes that
this is just an analytic continuation of (\ref{totalenergy}) in the masses:
\be
i\sqrt{8GM_{BH}}=1-4Gm_P
\ee
This suggests that the mapping function for two BHs will be some analytic
continuation of (\ref{Fuchs}) in the mass parameters $\mu_i,\mu_\infty$. We did
however not pursue this issue any further.

\section{Closed surfaces and the Gauss-Bonnet theorem}
In the previous sections we have seen how to accommodate particles and BHs in
open
spaces. But we could also decide to solve the Liouville equation on a closed
surface $\Sigma$. Because in the general solution to the Liouville equation we
still have an arbitrary analytic mapping function at our disposal, we expect
that a large class of Riemann surfaces are solutions to the Liouville equation
for $\Lambda<0$. There is however an important constraint on the possible
configurations of particles and BHs living on a closed Riemann surface with
constant negative curvature. Let's call $g$ the number of BHs, N the number of
BHVs and P the number of particles with mass $m_P$. The BH solutions have
hyperbolic monodromy and are therefore represented as handles on the surface.
The BHV solutions correspond to parabolic singularities (at infinity) and the
particles correspond to elliptic (or conical or Fuchsian) singularities on
$\Sigma$. The Gauss-Bonnet theorem tells us that the Euler characteristic can
be related to the intrinsic curvature of the Riemann surface:
\be
\chi(\Sigma)=\frac{1}{\pi}\int_{\Sigma}\sqrt{^{(2)}g}~^{(2)}R
\ee
Inserting the equation for the curvature (\ref{curvature}) in the above
equation and using the fact that $\chi(\Sigma)=2-2g$ gives:
\ba
2-2g&=&\frac{-1}{\pi}\int_{\Sigma}\nabla^2\phi\\
    &=&\int_{\Sigma}(\frac{-1}{2\pi}e^\phi+\sum_N\de^2(z-z_N)+4G\sum_P
m_P\de^2(z-z_P))
\ea
Because $\int e^\phi=\int\sqrt{^{(2)}g}={\bf Area}(\Sigma)$ we find finally:
\be
{\bf Area}(\Sigma)=2\pi(2g-2+N+4G\sum_P m_P)>0
\ee
So if there are no particles present, we would at least need 2 handles, or 1
handle and 1 parabolic singularity, or 3 parabolic singularities on a closed
surface of constant negative curvature. This is precisely the content of the
uniformization theorem on Riemann surfaces \cite{Matone}.

\section{de Sitter Space-Time: $\Lambda>0$}
\begin{figure}[t]
\centerline{\psfig{figure=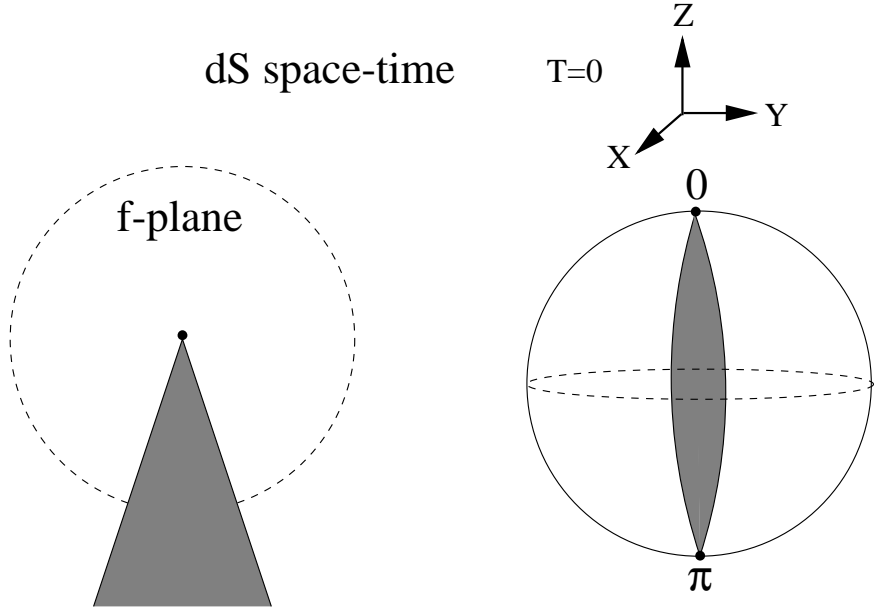,angle=-90,height=8cm}}
\caption{A particle located at the origin in dS space-time.}
\label{dS}
\end{figure}

In this section we want to consider briefly de Sitter space-time, i.e.
$\Lambda=\frac{1}{\ell^2}>0$. We assume again that the metric is of the general
form (\ref{lineelement}). The embedding in 3+1 dimensions is given by:
\ba
X&=&\ell\cosh\tau\sin\ta\cos\vfi\\
Y&=&\ell\cosh\tau\sin\ta\sin\vfi\\
Z&=&\ell\cosh\tau\cos\ta\\
T&=&\ell\sinh\tau
\ea
with:
\be
X^2+Y^2+Z^2-T^2=\ell^2
\ee
So at negative time this represents a contracting sphere of radius
$\ell\cosh\tau$, at $\tau=0$ the sphere has reached its minimal radius $\ell$
after
which it will start to expand.
The metric on de Sitter space is:
\ba
ds^2&=&-dT^2+dX^2+dY^2+dZ^2\\
    &=&\ell^2(-d\tau^2+\cosh^2\tau(d\ta^2+\sin^2\ta d\vfi^2))
\ea
After defining again $t=\tau\ell$, $S(t)=\ell\cosh(\frac{t}{\ell})$ and
$z=\tan\frac{\ta}{2}e^{i\vfi}$ this metric can be written as:
\be
ds^2=-dt^2+\frac{4S^2(t)}{(1+z\bz)^2}dzd\bz
\ee
The intrinsic and extrinsic curvature of the two surface is now given by:
\ba
^{(2)}R&=&\frac{2}{S^2}=\frac{2}{\ell^2\cosh^2(\frac{t}{\ell})}\\
K&=&\frac{2}{\ell}\tanh(\frac{t}{\ell})
\ea
This solves the contracted Gauss-Codazzi equation (or equivalently the
$00$-component of the Einstein equation):
\be
^{(2)}R+\ha K^2-\frac{2}{\ell^2}=0
\ee
If we allow for a more general conformal factor $e^{\phi(z,\bz)}$, then the
Einstein equation for this factor is:
\be
\dd\db\phi+\ha e^\phi=-4\pi G\sqrt{^{(2)}g}T^{00}\label{liouville2}
\ee
The Einstein equation for the scale factor $S(t)$ is precisely solved by
$S=\ell\cosh(\frac{t}{\ell})$ if we assume time symmetry at $t=0$. The general
solution for the slightly altered Liouville equation (\ref{liouville2}) is:
\be
e^\phi=\frac{4\dd g~\db\bar{g}}{(1+g\bar{g})^2}
\ee
If we add one particle at $R^2\equiv g\bar{g}=0$ we can simply use the Ansatz
$g(z)=(z)^\mu$ again and find: $\mu=1-4Gm$. After a coordinate transformation
to static coordinates this turns out to be the solution found by Deser and
Jackiw in \cite{DJ}. In contrast with AdS space the metric does not blow up at
$g\bar{g}=1$. If we calculate the distance from $R=0$ to $R=\infty$ at a
constant time we find:
\be
\Delta x=S(t)\int_0^\infty dR~\frac{2}{(1+R^2)}=\pi S(t)
\ee
This implies that the singularity at $R=\infty$ is actually located at the
opposite site of the sphere with radius $S(t)$, i.e. it is antipodal: see
figure (\ref{dS}).
Adding more particles is only possible in even numbers therefore. In the case
of $2N$ particles we are faced with a Riemann-Hilbert problem with $2N$
SU(2)-monodromies that cancel pairwise. It is obvious that the mapping function
$g(z)$ will be very complicated in this case (although solvable in principle).

\section{Discussion}
In this paper we studied cosmologies of expanding or contracting matter in de
Sitter, anti-de Sitter and Minkowski space-time. The AdS cosmologies correspond
to the ones described by Steif and Brill \cite{Steif,Brill,Brill2}, i.e. they
are time symmetric at $t=0$. We found that the metric can be completely
described in terms of a single factor composed of the product of a
time-dependent scale $S(t)$ and space-dependent conformal factor
$e^{\phi(x,y)}$.
$S(t)$ describes the scale of the spatial surface at a time $t$. Amazing, and
typical for gravity in 2+1 dimensions is that this scalefactor is not
influenced by the matter content of the universe, i.e. it is the same for the
vacuum solution and for a universe filled with mass. This is due to the
topological nature of the interactions in 2+1 dimensions: there is no `real
force' acting on the masses. In other words, there are no gravitons to carry
the interaction from one object to the other. The effect of the mass
distribution is described the conformal factor $e^\phi$. The Einstein equations
for this conformal factor translate into the Liouville equation for all three
possibilities of the cosmological constant (dS space has however an extra minus
sign in the equation). The solution of this Liouville equation is completely
determined in terms of a single analytic function $f(z)$. So once we determine
this mapping function for a matter distribution we know the complete metric and
its time evolution. In 2+1 D AdS space-time there exist three kinds of massive
object:  point particles, black holes and black hole vacuum solutions. Note
however that the black hole vacuum solution is not vacuum AdS space but an
intermediate state between a particle and a black hole (a particle with mass
$1/4G$ or a BH with zero mass). It happens that when we add these objects to
AdS space the mapping function becomes multivalued. It has elliptic monodromy
for particles, hyperbolic monodromy for BHs and parabolic monodromy for BHVs.
So to find $f(z)$ for a combination of these three would result in
Riemann-Hilbert problem with three types of monodromy \cite{Welling,Bellini}.
We found explicit solutions in the following cases:
\ba
\Lambda=0,\frac{-1}{\ell^2}&~~~~~~~~~~~&\mbox{1 particle, 2 particles, 1 BH, 1
BHV}\\
\Lambda=\frac{+1}{\ell^2}&~~~~~~~~~~~&\mbox{2 particles}
\ea
Notice that a 1 particle solution does not exist in dS space as a particle
always has an antipode. The explicit solutions for AdS and Minkowski space-time
represent open spatial surfaces (an open surface is defined as a surface with
infinite area). In the case of dS space-time the spatial slice is always
closed. In the last section we studied the possibility of closed surfaces in
AdS and Minkowski space. Using the Gauss-Bonnet theorem we derive a simple
inequality that determines which combinations of particles, BHs and BHVs can
live in a closed space. This inequality is consistent with the uniformization
theorem for Riemann surfaces. Finally we would like to comment on a possible
application of the linearly expanding or contracting solution in the case of
vanishing cosmological constant. It is well known that a serious complication
in the calculation of a scattering amplitude using perturbation theory is that
the
particles are never free of `gravitational interaction' in 2+1 D gravity. The
interaction
persists all the way to infinity because asymptotically the metric of a N
particle solution is a cone and not Minkowski space. Actually the particles
will be described for $t\ra\infty$ by the metric (\ref{lineelement}) (with
$S(t)\sim t$). The mapping function will be very complicated in the
general case of N particles. Say however that we can define a wavefunction
$\Psi_{\bf in}(z_1,...z_N)$ that describes this asymptotic state. On the
Poicar\'e disc we have the angle at which the particle moves radially outward
and the distance to the origin as a measure for its momentum. Then a possible
definition for the scattering matrix is:
\be
<\Psi_{\bf out}(z_1,...,z_N)|\Psi_{\bf in}(z_1,...,z_N)>
\ee

\section{Acknowledgements}
I would like to thank G. 't Hooft and E. Verlinde
for valuable suggestions and discussions.


\begin{thebibliography}{99}
\bibitem{gravitation} Misner C W, Thorne K S and Wheeler J A 1973 {\em
Gravitation}
(San Francisco: W H  Freeman and Company)
\bibitem{DJH} Deser S, Jackiw R and 't Hooft G 1984
{\em Ann Phys } {\bf 152} 220
\bibitem{DJ} Deser S and Jackiw R 1984 {\em Ann  Phys } {\bf 153} 405
\bibitem{tHooft}  't Hooft G 1992 {\em Class  Quant  Grav } {\bf 9}  1335
\bibitem{witten} Achucarro A and Townsend P 1986 {\em Phys  Lett }
                  B{\bf 180} 85
\bibitem{Witten} Witten E 1988 {\em Nucl  Phys } B{\bf 331} 46
\bibitem{Welling} Welling M 1996 {\em Class Quant Grav} {\bf 13} 653
\bibitem{Bellini} Bellini A, Ciafaloni M and Valtancoli P 1995
{\em Phys Lett} {\bf B357} 532, 1996 {\em Nucl Phys} {\bf B462} 453
\bibitem{Ciaf} Ciafaloni M and Valtancoli P 1997
{\em Class Quant Grav} {\bf 14} 955
\bibitem{BTZ}  Ban\~ados M, Teitelboim C and Zanelli J 1992 {\em Phys Rev Lett}
{\bf 69} 1849
\bibitem{Steif} Steif A R 1996 {\em Phys Rev} {\bf D53} 5521
\bibitem{Brill} Brill D R 1996 {\em Phys Rev} {\bf D53} 4133
\bibitem{Brill2} Brill D R 1996 {\em Geometry of Black Holes
and Multi-Black-Holes in 2+1 dimensions} {\bf gr-qc/9607026}
\bibitem{Sachs} Sachs R K 1964 {\em Gravitational radiation} in {\em
Relativity, Groups and Topology (Les Houches 1963)} eds DeWitt C and DeWitt B
(New York: Gordon and Breach)
\bibitem{Clement} Cl\'ement G 1994 {\em Phys Rev} {\bf D 50} 7119
\bibitem{Matone} Matone M (1995) {\em Int  J  Mod  Phys } A{\bf 10} 289
\bibitem{Plemelj} Plemelj J 1964 {\em Problems in the sense of Riemann and
Klein} (London: John Wiley)
\bibitem{Sato} Sato M, Miwa T and Jimbo M 1979 in {\em Complex Analysis,
Microcalculus, and Relativistic Quantum Theory (Lecture Notes in Physics, vol
120)} ed Lagolnitzer D  (Berlin: Springer)
\bibitem{Chud} Chudnovsky D V 1979 {\em Nonlinear Evolution Equations and
Dynamical Systems (Lecture Notes in Physics, vol 120)} ed Boiti M Pempinelli F
and Soliani G (Berlin: Springer)
\bibitem{complex} Sansone G and Gerretsen 1969 {\em Lectures on the theory of
functions
of a complex variable II} (Groningen: Wolters- Noordhoff)
\bibitem{Hille} Hille E 1976 {\em Ordinary differential equations in the
complex domain} (New York: John Wiley \& Sons)
\bibitem{Bieberbach} Bieberbach L 1953 {\em Theory der gew\"onlichen
differentialgleichungen} (Berlin: Springer)

\end{thebibliography}
\end{document}